\documentclass[showpacs,preprintnumbers,amsmath,amssymb,superscriptaddress]{revtex4}

\usepackage{mathrsfs}
\usepackage{graphicx}
\usepackage{dcolumn}
\usepackage{bm}

\begin{document}
%\begin{CJK*}{GB}{gbsn}

\preprint{Phys. Rev. E \textbf{77}, 032103 (2008).}

\title{Emergence of synchronization induced by the interplay between two prisoner's dilemma games with volunteering in small-world networks}

\author{Yong Chen}
\altaffiliation{Email address: \tt {ychen@lzu.edu.cn}} \affiliation{Research Center
for Science, Xi'an Jiaotong University, Xi'an 710049, China} \affiliation{Institute
of Theoretical Physics, Lanzhou University, Lanzhou 730000, China}

\author{Shao-Meng Qin}
%\altaffiliation{Email address: \tt {qsminside@gmail.com}}
\affiliation{Institute of Theoretical Physics, Lanzhou University, Lanzhou 730000,
China}

\author{Lianchun Yu}
\affiliation{Institute of Theoretical Physics, Lanzhou University, Lanzhou 730000,
China}

\author{Shengli Zhang}
\affiliation{Research Center for Science, Xi'an Jiaotong University, Xi'an 710049,
China}

\date{\today}

\begin{abstract}
We studied synchronization between prisoner's dilemma games with voluntary
participation in two Newman-Watts small-world networks. It was found that there are
three kinds of synchronization: partial phase synchronization, total phase
synchronization and complete synchronization, for varied coupling factors. Besides,
two games can reach complete synchronization for the large enough coupling factor.
We also discussed the effect of coupling factor on the amplitude of oscillation of
$cooperator$ density.
%\textbf{Keywords:} synchronization, Prisoner's Dilemma Game, complex networks
\end{abstract}

\pacs{02.50.Le,87.23.Kg,87.23.Ge,05.45.Xt}
% Dynamics of evolution
% Dynamics of social systems
% Synchronization, coupled oscillators
% Decision theory and game theory
% Networks and genealogical tree

\maketitle

%\end{CJK*}

There has been a long history of studying game theory. Some restricted version of
the Nash equilibrium concept as early as $1838$ was used by the French economist
Augustin Cournot to solve the quantity choice problem under duopoly \cite{duopoly}.
Since the comprehensive seminal book of Neumann and Morgenstern, game theory has
become a powerful framework to investigate evolutionary fate of individual traits
under differing competition~\cite{Economic}. In recent years, it has been applied
successfully to problems in biology~\cite{biology}, psychology, computer science,
operation research, political science, military strategies, economics, and so
on~\cite{book1,book2,sazbo06}.

The prisoner's dilemma game (PDG) stands as a paradigm of a system which is capable
of displaying both cooperative and competitive behaviors through pairwise
interactions. After PDG was first applied by Axelrod on a lattice~\cite{Axelrod},
spatial prisoners' dilemma games (SPDGs) have been studied in various kinds of
network models. In the general PDG, each of two players has two strategies,
\textit{cooperator} ($C$) or \textit{defector} ($D$). If both of them choose the
$C(D)$, the player will get payoff $R$($P$). When the $D$ betray the $C$, the $D$
will win the income $T$ and the $C$ gets $S$. Four elements satisfy the order
ranking $T>R>P>S$ and usually the additional constraint $(T+S) < 2R$ in repeated
interaction. So, the mutual cooperation leads to the highest return for the
community and defection is the optimal decision regardless of the other player. In
SPDG, the players are located on the nodes. Each player updates his strategy at
discrete time steps and has a probability of mimicking his neighborhood strategies.

Szab\'o \textit{et al.}~\cite{Szabo} developed SPDG with voluntary participation, in
which players take one of the three strategies, $C$, $D$, or \textit{loner}($L$). In
the traditional PDG, every player participates in this game compulsorily. However,
players might drop off risky social enterprise. They do not participate in the game
temporarily and earn a smaller payoff $\sigma$ ($0<\sigma<R$) on their individual
efforts. $L$ is better than a pair of $D$, but is less than two $C$. If one of the
two players chooses $L$, the other player is forced to choose $L$. The payoff is
determined by the matrix in table~\ref{table-1}.

\begin{table}[h]
\caption{The payoff matrix of spatial prisoners' dilemma game with voluntary
participation.}
\begin{ruledtabular}
\begin{tabular}{c|c c c}
 player1$\setminus$player2 & \textit{C}   &\textit{D} & \textit{L} \\ \hline
  \textit{C} & $ R \setminus R$ & $S \setminus T$ & $\sigma \setminus \sigma$\\
  \textit{D} & $ T \setminus S$ & $P \setminus P$ & $\sigma \setminus \sigma$\\
  \textit{L} & $ \sigma \setminus \sigma$ & $\sigma \setminus \sigma$ & $\sigma \setminus \sigma$\\
\end{tabular}
\end{ruledtabular}
\label{table-1}
\end{table}

The purpose of this paper is to describe some interesting peculiarities between two
local stock markets or two stocks in one market by the model of PDG with voluntary
participation and Newman-Watts small-world (NWSW) networks~\cite{NWSW}. The stock
trading can be regarded as a prisoner's dilemma game. Shareholders gain if everyone
chooses to cooperate. If any large shareholders choose to sell, the remaining
shareholders will likely lose. NWSW can mimic the properties of social
networks~\cite{NWSW}.

There are economic phenomena that one local stock market follows the fluctuations of
other stock markets and occasional synchronous events happen among the individual
stocks. The correlation between different stock markets has a pivotal role in
value-at-risk (VaR) measures, optimal portfolio weights, hedge rations and so
on~\cite{jbf}. The synchronous events among the individual stocks also important to
the stock market asymmetry~\cite{jstat}. In ref.~\cite{jstat}, R. Donangelo believes
that the synchronous events are caused by \textit{fear factor}. However, why
shareholders do not feel exciting together when there are big chances to obtain the
profit? Generally speaking, this phenomenon is induced by the interactions among
shareholders of different stocks and  in different domestic financial markets.  In
our opinion, the synchronization is caused by the conformity. It means that when we
find ourselves in the minority in a groups, we could change our decision to avoid
uncomfortable caused by that situation. From the social psychology point of view, we
understand the fact that conformity is pervasive applicable~\cite{conformity}.

Our model is defined as follows. We consider two NWSW networks with the same size
and the same rewired bound $p$ but different detailed initial structures. NWSW
network is obtained in practice by the following procedures.
\begin{enumerate}
\item[(a)] Starting with a 2-dimensional lattice network with periodic boundary, each player is located on the node with four neighbors;
\item[(b)] The long range links are randomly rewired with a certain fraction $p$.
\end{enumerate}
Every node in networks is an agent in the game. Each agent is a pure strategist and
can only take one of the three strategies ($C$,$D$,$L$). In each simulation time
step, all agents play PDG with their neighborhoods. The parameters in the payoff
matrix are $R=1$, $S=P=0$, $T=b$ ($1.0<b<2.0$), and $\sigma=0.3$, where $b$ is
regarded as temptation. At next time step, $\emph{i}$th player changes its strategy
by following one of its neighborhoods that is selected with equal probability. The
probability of this change is defined as,
\begin{equation}
W = \frac{1}{1+\exp \left[ -(E_i-E_j)/\kappa \right] }, \label{eq1}
\end{equation}
where $E_i$, $E_j$ are the $\emph{i}$th and $\emph{j}$th players' total payoffs at
the previous round. $\kappa=0.1$ denotes the noise to permit irrational choices. The
payoffs of the players will not be counted in the next round. In the case of
$\kappa=0$, the strategy of player $\emph{j}$ is always adopted provided $E_j>E_i$.
We found that $\kappa$ does not have obvious effect in our model if it is within
realistic limit~\cite{Szabo}. So, in our model, we only have two parameters, the
rewired fraction $p$ and the temptation $b$.

In order to model the conformity we make an assumption that agents could be affected
by the other network. The point of our assumption that the network could not affect
its agents is that people usually cannot obtain the perfect information from their
group and their neighborhoods will affect their judgement of global situation. We
define a coupling factor $F$ to actualize this assumption. Now, we begin by
describing two interplayed PDGs in NWSW networks with the following modifications:

\begin{enumerate}
\item[(a)] At each step, each agent chooses strategy $C$ with a probability $F\times C_{other}$, where $C_{other}$ is the $C$ density in the other network.
\item[(b)] The agent will search for a better strategy according to the rule mentioned above.
\end{enumerate}

Clearly, the larger $F$, the easier one agent is influenced by the agents in the
other network. For $F=0$, all agents play PDGs in the network without interplay. A
similar model that the players in PDG are influenced by external constraints has
been studied by Szab\'{o}~\cite{1095}.

\begin{figure}[h]
\begin{center}
\includegraphics[width=0.5\textwidth]{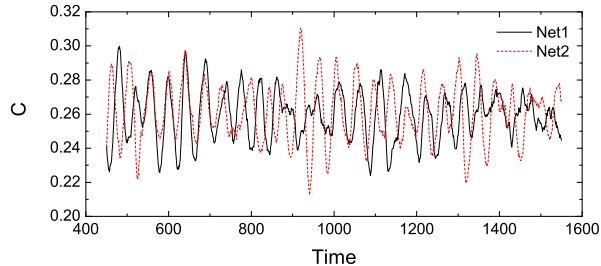}
\caption{(Color online). The oscillations of $C$ strategy in two different NWSW
networks. Parameters are $b=1.6$ and $p=0.01$. } \label{fig1}
\end{center}
\end{figure}

Works by G. Szab\'o, C. Hauert and Z.-X. Wu reported a comprehensive study of this
model in the case of $F=0$~\cite{Szabo,CH,wuzhixi}. One of the most important
characters of this model is the persistent global periodical oscillations of three
different strategies. If most of the players select $C$, $D$ will be more
profitable; however, when $D$ is dominative, $L$ will make a steady income; after
$L$ alleviate the threat of $D$, $C$ attracts the $L$ to converse. So, three
strategies implement a rock-scissors-paper-type cyclic dominance. However, the
periodical oscillations of strategies in two different NWSW networks do not show the
same amplitude and phase (see Fig.~\ref{fig1}). The reason for this difference is
that two network structures are not exactly identical and the evolution of SPDG is
random.

To measure the differences (or synchronization) between PDGs in two NWSW networks,
one conspicuous parameter is $\Delta C$. It is defined as,
\begin{equation}
    \Delta C = \frac{1}{N} \sum_{t=0}^N \left|c_1(t)-c_2(t)\right|,
    \label{define1}
\end{equation}
where $c_1(t)$ and $c_2(t)$ are the $C$ density at step $t$. Clearly, $\Delta C=0$
means complete synchronization state. Because there exists the global period
oscillation of three strategies in this model, one can define the phase of strategy
$C$ in every time step. In this paper, we use definition \textbf{B} in
Ref.~\cite{phase}. IIn this paper, we use the crossing of mean $C$ density as the
beginning of the cycle. To study the phase of oscillation, one sets $\Delta \varphi$
as
\begin{equation}
    \Delta \varphi = \frac{1}{N} \sum_{t=0}^N \left|\varphi_1(t)-\varphi_2(t)\right|,
    \label{define2}
\end{equation}
where $\varphi_1(t)$ and $\varphi_2(t)$ are the phases of strategy $C$ at step $t$
in two networks. $\Delta \varphi=0$ indicates there are no phase differences between
PDGs in two NWSW networks. The maximal phase difference is $\pi$ and $\Delta \varphi
= \pi/2$ means no phase synchronization. However, when two networks reach a
synchronization state with constant phase difference $\varphi=\pi/2$, we cannot
distinguish synchronization from no synchronization. To avoid this puzzle, another
parameter $Q$ defined by Kuramoto~\cite{Kuramoto} is introduced
\begin{equation}
        Q = \left| \frac{1}{N} \sum_{t=0}^Ne^{i \left[ \varphi_1(t) - \varphi_2(t) \right]} \right|.
    \label{define3}
\end{equation}

In the case of phase synchronization, all vectors in complex space of $e^{i\left[
\varphi_1(t)-\varphi_2(t) \right]}$ have the same direction, and $Q$ is close to
$1$. $Q=1 $ means total synchronization and $0<Q<1$ means partial synchronization,
while for $Q=0$ there is no synchronization at all.

\begin{figure}[h]
\begin{center}
\includegraphics[width=1\textwidth]{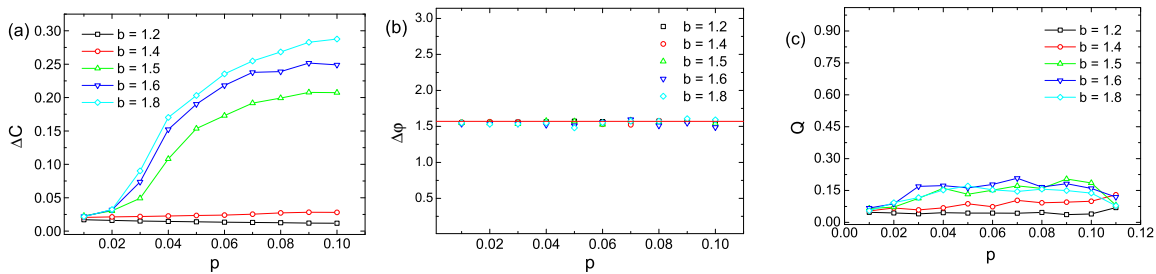}
\caption{(Color online). The differences of $C$ strategy between two PDGs without
interplay ($F=0$). The red solid line in (b) is $\Delta \varphi = \pi /2$. }
\label{fig2}
\end{center}
\end{figure}

Fig.~\ref{fig2} shows the three synchronization parameters depended on $p$ for PDGs
in two NWSW networks with various $b$ for $F=0$. Considering that strategy density
in this model is periodical oscillation, after initial transient state, we recorded
$25000$ steps as a sampling to calculate these parameters. And, the behaviors of
this model will be affected by the size of network and random seed. All simulations
in this paper are performed in networks with $200 \times 200$ players and random
initial states with an equal fraction of three strategies. All results presented in
this paper are the average of $20$ trials with different random seed. Obviously,
$\Delta \varphi$ fluctuates around $\pi /2$, and $Q$ is close to $0$. This clearly
demonstrates that the synchronization state does not exist. For $p \leq 0.2$ or $b <
1.4$, $\Delta C$ is very small. For $1.5 \le b \le 1.9$, $\Delta C$ increases
monotonically with $p$. As to $b> 1.9$, the oscillations of strategies become large
enough to inevitably lead to the extinction of one strategy. The amplitude of
oscillation depends on $p$ and temptation $b$ are conformed by Szab\'o and
Hauert~\cite{Szabo}.

\begin{figure*}
\begin{center}
\includegraphics[width=1\textwidth]{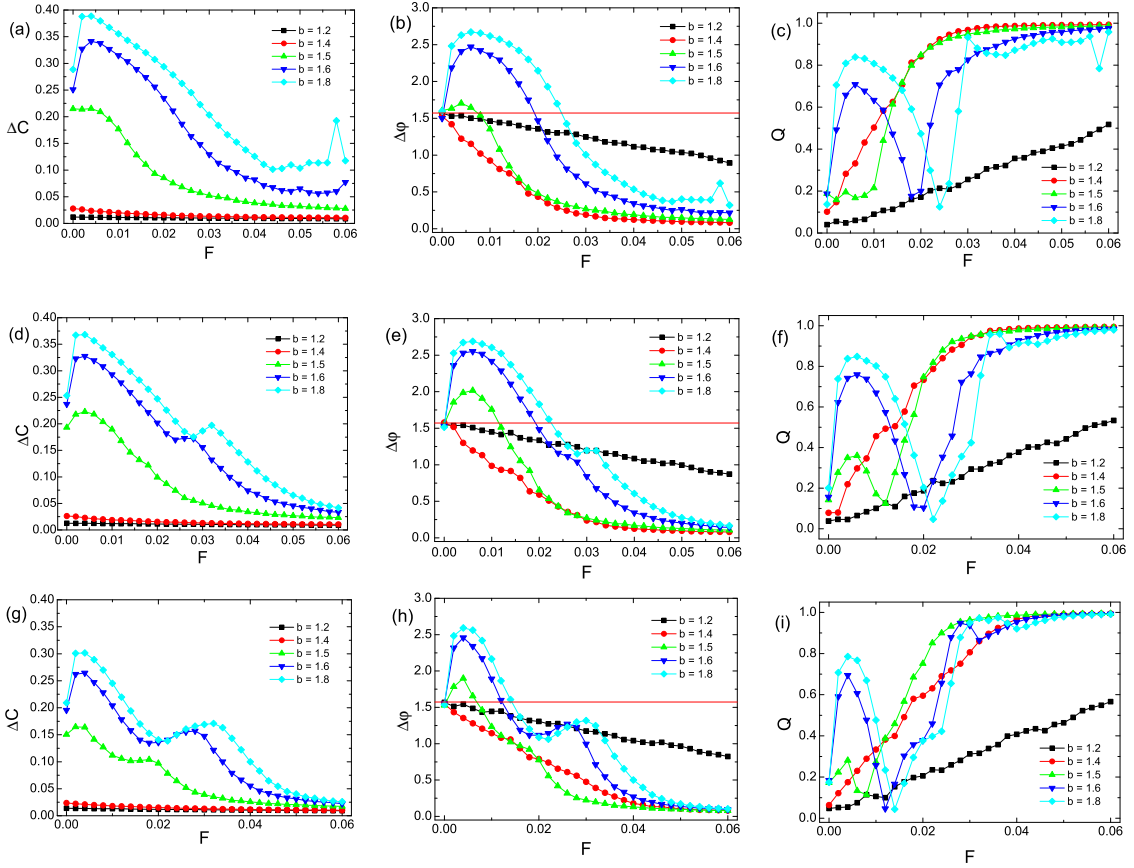}
\caption{\label{fig3} (Color online). From top to the bottom $p=0.1$, $0.075$, and
$0.05$. The red solid line is $\Delta \varphi = \pi /2$. }
\end{center}
\end{figure*}

In Fig.~\ref{fig3}, we plot how three parameters  $\Delta C$, $\Delta \varphi$, and
$Q$ vary with the coupling factor $F$. Clearly, it shows that synchronization state
exists when $F$ is large enough. For small $b$ $b \le 1.3$, $\Delta C$ is very close
to $0$ and it looks like an identical synchronization with different phases. For
larger $b$, by enhancing $F$ enough, the networks should reach a synchronization
with $\Delta \varphi \approx 0$ and $Q \approx 1$.

It is interesting that three parameters do not always increase or decrease
monotonously with $F$. For example, when $b=1.8$ and $p=0.05$, all three parameters
increase up to the first peak at $F = 0.004$ which indicates two networks are not
independent and there is a phase discrepancy between them. It means that a partial
phase synchronization emerges between two networks. The largest phase discrepancy in
this stage is close to $\pi$. As $F$ increases near to $0.014$, three parameters
decrease to their level at $F=0$ and the partial phase synchronization is broken.
There exists a gap in the Fig.~\ref{fig3}(c,f,i) at $F=0.014$. This gap (the first
gap) is the boundary between partial and total phase synchronization. Till the $F$
reaches the range $0.03<F<0.04$, two networks become correlative once again to reach
a total phase synchronization. In this range, $\Delta C$ and $\Delta \varphi$ have
an abnormal bump. It denotes an augmenting phase discrepancy in phase
synchronization. After this slight improvement of coupling, two networks reach
complete synchronization. However, we find that $Q$ has a small gap (the second gap)
at $F=0.04$. This gap becomes the boundary between complete and phase
synchronization. As the above discussion, an interesting phenomenon is that the
effect of $F$ is not continuous. In order to achieve a new synchronization, the old
one should be broken down first.

$\Delta C$ and $\Delta \varphi$ increase with the structural parameters $p$ and $b$
in partial synchronization stage. $p$ and $b$ are important to the position of the
first gap. The larger $p$ and $b$, the larger $F$ for the first gap position.
Moreover, $Q$ is closer to $1$ for larger $p$ and $b$ with the same $F$ on the
partial synchronization. It can be seen from the Fig.~\ref{fig3} that the increase
of $p$ makes the process of synchronization more difficulty. The abnormal bump of
$\Delta C$ and $\Delta \varphi$ and the second gap disappear gradually on larger $p$
and smaller $b$. Three different kinds of synchronization can only be observed in
the case of large $b$ and small $p$. Since the large enough amplitude of oscillation
inevitably leads to the extinction of one strategy, the synchronization will become
unstable for large $p$ and $b$.

\begin{figure}[h]
\begin{center}
\includegraphics[width=0.5\textwidth]{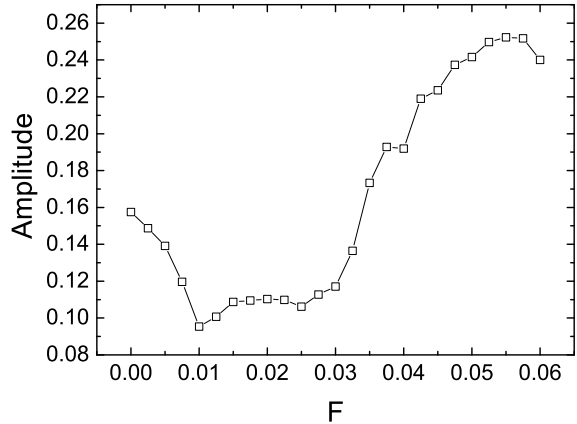}
\caption{\label{fig4} The amplitude of the oscillation of $C$ density for varied
coupling $F$. Parameters are $b=1.6$ and $p=0.05$.}
\end{center}
\end{figure}

Fig.~\ref{fig4} presents how the coupling $F$ affects the amplitude of oscillation
of $C$ density. It was found that at the partial phase synchronization stage the
amplitude decreases with the coupling factor $F$. In the regime where the $\Delta
\varphi$ has the abnormal bump, the amplitude increases slightly. Then, the
amplitude increases monotonously with larger $F$. Comparing the amplitude with
$\Delta \varphi$, it is easy to find that the behaviors of $\Delta \varphi$ and
amplitude with $F$ are contrary. It is conjectured that there is a positive feedback
for the interplay between games in two networks.

In this work, we discussed the effect of the interplay between two prisoner's
dilemma games with volunteering in NWSW networks. By defining the coupling factor
$F$ between two different networks based on the conformity psychology, it was found
that the large enough $F$ will lead to synchronization between two networks. We
concluded that this model captures the synchronization characteristic in stock
markets. To measure the detailed information of synchronization, we introduced three
parameters, $\Delta \varphi$, $\Delta C$, and $Q$. It shows that there are three
different kinds of synchronization for different $F$ from our extensive simulations.
The network structure $p$ and the temptation $b$ play a very important role on the
synchronization between two networks.

\bigskip

This work was supported by the National Natural Science Foundation of China under
Grant No. $10305005$ and by the Special Fund for Doctor Programs at Lanzhou
University.

\end{document}